\documentclass[aps,prxquantum,twocolumn,groupedaddress]{revtex4-1}
\usepackage{graphicx}
\usepackage{dcolumn}
\usepackage{bm}
\usepackage{epstopdf}
\usepackage{float}
\usepackage{amsmath}
\usepackage{xcolor}
\usepackage{comment}
\usepackage[normalem]{ulem}
\graphicspath{{./images/}}

\begin{document}
\title{Topological Frenkel Exciton-Polaritons in One-Dimensional Lattices of Strongly Coupled Cavities}
\author{J. Andr\'es Rojas-S\'anchez}
\author{Yesenia A. Garc\'ia Jomaso}
\author{Brenda Vargas}
\author{David Ley Dominguez}
\author{C\'esar L. Ordo\~nez-Romero}
\author{Hugo A. Lara-Garc\'ia}
\author{Arturo Camacho-Guardian}
\email{acamacho@fisica.unam.mx}
\author{Giuseppe Pirruccio}
\email{pirruccio@fisica.unam.mx}
\affiliation{Instituto de F\'isica, Universidad Nacional Aut\'onoma de M\'exico, Apartado Postal 20-364, Ciudad de M\'exico C.P. 01000, Mexico}

\date{\today}
\begin{abstract}
Frenkel polaritons, hybrid light-matter quasiparticles, offer promises for the designing of new opto-electronic devices. However, their technological implementations are hindered by sensitivity to imperfections. Topology has raised as a way to circumvent defects and fabrication limitations. Here, we propose a lattice of cavities to realize the one-dimensional Su-Schrieffer-Heeger model (SSH) for topological Frenkel polaritons. By engineering the configuration of the cavities we demonstrate that the SSH topological and trivial phases can be accessed, which we unravel by complementary classical and quantum theories. We demonstrate the polariton edge state robustness under defects and the broadening of the photon and exciton lines. Our findings propose a realistic yet simple experimental setup to realize topological room temperature polaritons.

\end{abstract}

\maketitle

\section{Introduction}
Frenkel excitons have emerged as a successful platform  to realize strongly hybridized phases of light and matter at room temperature~\cite{Lidzey1998,kena2010room}. Experimental breakthroughs have demonstrated the ability to produce many-body phases such as Bose-Einstein condensation~\cite{cookson2017yellow,plumhof2014room,scafirimuto2018room,betzold2019coherence}, superfluidity~\cite{lerario2017room}, and a variety of effects resulting from exciton-polariton ~\cite{daskalakis2014nonlinear,yagafarov2020mechanisms} and plasmon-exciton-polariton interactions,\cite{PhysRevLett.101.116801,torma14,Ramezani:19,rivas,jana,rivas2} including lasing~\cite{wei2019low,ballarini2014polariton,mazza2013microscopic}, polariton parametric emission \cite{parametric} and oscillation \cite{oscillation,Kuznetsov:20}, among others. The flexibility of these systems permits the polaritonic control of the internal energy levels~\cite{eizner2019inverting,stranius2018selective}, has opened up the field of polaritonic chemistry~\cite{Keeling2020,Sanchex2022,Liu2020,du2022catalysis}, and encouraged studies beyond the quasiparticle approach of polaritons~\cite{garcia2022fate}. 
The ultimate control of strongly coupled light-matter excitations, paired up with the emerging field of topological photonics, paved the way to the advent of topological polaritonics ~\cite{ozawa2019topological,lu2014topological,karzig2015topological}. This may boost technological applications in quantum optical circuits~\cite{blanco2019topological}, non-linear light~\cite{smirnova2020nonlinear}, chiral and topological lasers~\cite{harari2018topological,bandres2018topological,jimenez2017chiral}, and in general where high fabrication precision is challenging to be reached.

The advances in topological photonics and polaritonics, include breakthrough experiments and theories in the context of cavity- and circuit-QED systems~\cite{schmidt2013circuit,qi2018simulation,mei2015simulation,Owens2018,Cho2018}, ring resonator arrays~\cite{lin2018three,hafezi2013imaging,mittal2014topologically,mittal2016measurement}, photonic cystals~\cite{wang2020topological,lu2016symmetry,wu2015scheme,skirlo2015experimental,raghu2008analogs,Malkova,PhysRevLett.112.107403}, microwaves~\cite{kuhl1998microwave,hu2015measurement,cheng2016robust,anderson2016engineering}, and metamaterials \cite{poli,PhysRevB.97.220301,microlaser,mechanics,menon} in which many intriguing topological phases have been realized exploiting the light-matter coupling. Perhaps, the canonical one-dimensional (1D) model with non-trivial topological properties is the so-called Su-Schrieffer-Heeger (SSH) chain. SSH models have already been realized and studied in many systems including plasmonic chains~\cite{bleckmann2017spectral,kruk2017edge,ling2015topological,poddubny2014topological,slobozhanyuk2015subwavelength}, waveguide QED~\cite{kim2021quantum}, radiative heat transfer \cite{biehs} and polaritons~\cite{solnyshkov2016kibble,st2017lasing,parto2018edge,kozin2018topological,downing2019topological,su2021optical,dusel2021room}. Topological edge states provides an efficient way to create localized polaritonic modes which are protected by their bulk environment. Room temperature topological systems are of particular interest because their robustness against fabrication imperfections may lead to next-generation polariton-based technologies. 
\begin{figure}[h]
\centering
\includegraphics[width=\columnwidth]{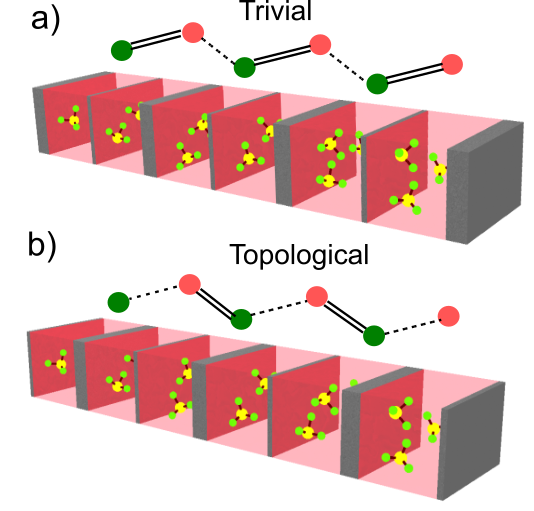}
\caption{Schematic representation of the array of $2N$ stacked cavities filled with an organic material  and its analogy to the 1D SSH model. Each cavity supports Frenkel exciton-polaritons. The different couplings found in the SSH model are realized by alternating the two mirror's width separating each active layer. (a) Trivial and (b) Topological configuration.}
\label{Fig1} 
\end{figure}

Here, we theoretically propose a room-temperature setup for the realization of the SSH model with Frenkel exciton-polaritons in a one-dimensional lattice of stacked nanocavities. We demonstrate that by alternating the width of the mirrors it is possible to obtain both trivial and non-trivial topological polariton phases. For this, we employ a dual approach based on the transfer matrix method (TMM) combined with a tight-binding model for a chain of exciton-polaritons. The TMM, for the appropriate configuration unveils the emergence of polariton edge states with localized electric field around the edges of the array. Concomitantly, the reflectance spectrum of the stack is found to closely resemble that of an isolated cavity. The correspondence of these branches with the topologically-protected states of the SSH model is demonstrated by means of the tight binding formalism for exciton-polaritons. Our twofold approach is general and provides a comprehensive tool that enables a deep understanding of the fundamental aspects of stacks of strongly coupled cavities.  In addition, we discuss the experimental implementation of our proposal and its robustness against typical fabrication limitations. Even though we mainly deal with Frenkel polaritons, our formalism is applicable to exciton-polaritons in inorganic materials. This is demonstrated in the last section where we consider homogeneously broadened excitons lacking vibronic coupling. 

Our proposal can be implemented in a wide family of organic polaritons at room temperature and provides therefore a valuable guide for future experiments and theories. An additional value of our proposal stems from the facile and cheap fabrication process combined with a simple and scalable design.

\section{System}
\label{System}
\begin{figure}[h]
\centering
\includegraphics[width=\columnwidth]{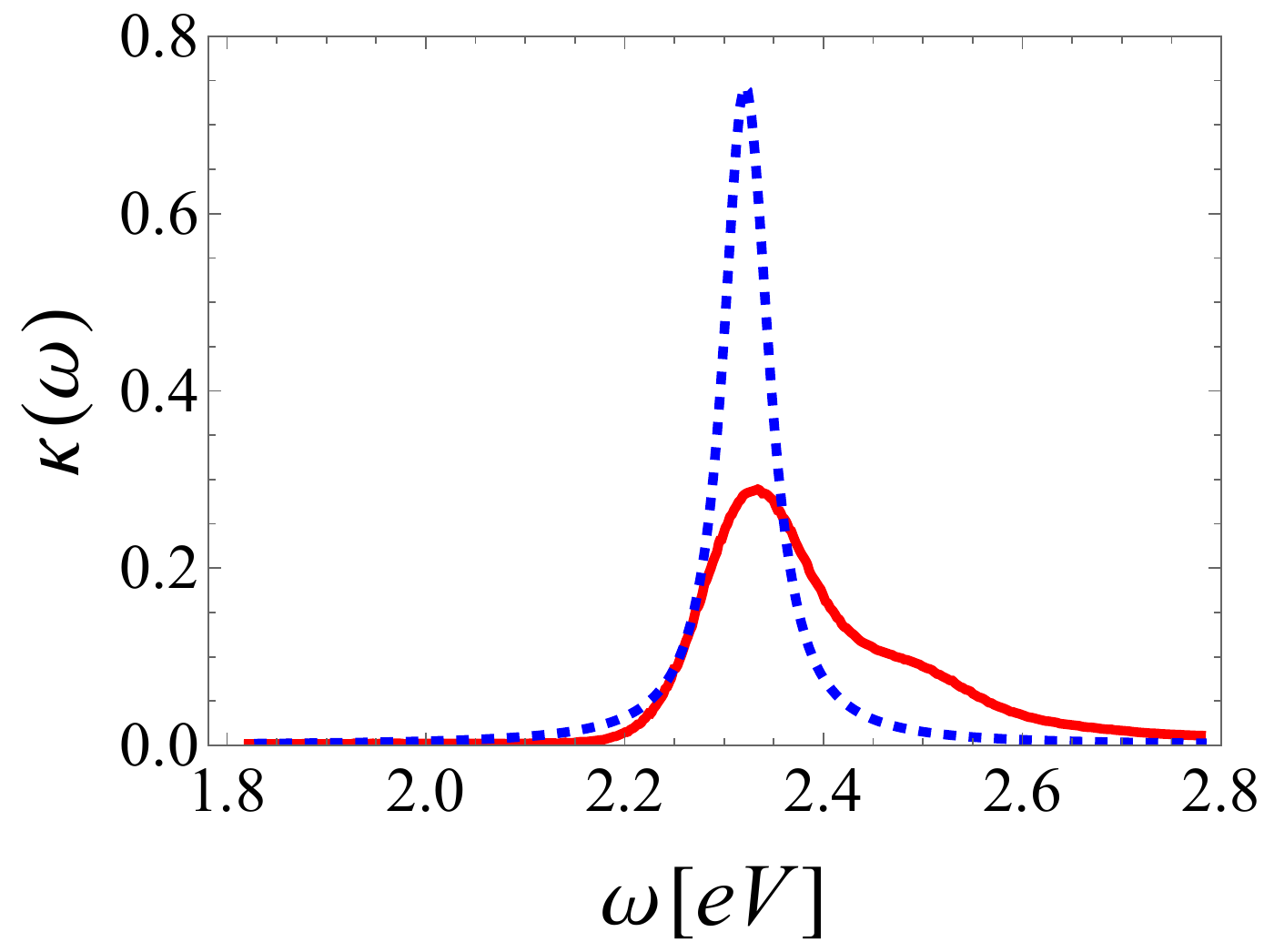}
\caption{Imaginary part of the refractive index. Solid red curve represents the spectrum of a typical organic molecule with a principal peak at $\omega_X\approx 2.32\text{eV}$ and a second vibron mode around $\omega_S\approx 2.5\text{eV}.$ Dashed blue curve represents an ideal exciton with an oscillatory strength of $2\Omega=0.33\text{eV}$ peaked at $\omega_X=2.32\text{eV}$ and an exciton linewidth of $\gamma_X=0.025\text{eV}$  }
\label{FigErB}
\end{figure}

Our system consists of an array of $2N$ stacked nanocavities, as illustrated in Fig.~\ref{Fig1}. All cavities are loaded with a polymer matrix mixed with a highly concentrated organic molecule. In Fig.~\ref{FigErB} we show the imaginary part $\kappa(\omega)$ of the refractive index for a generic organic molecule. It is formed by a principal electron transition, associated to the zero-phonon exciton line, strongly coupled to a vibronic sideband. In a typical organic molecule at room temperature, the vibronic shoulder is slightly detuned from the mean peak yet overlaps with it giving rise to a continuum of material excitations that cannot be disentangled. The cavity length, $L_c$, common to all cavities, is such that the fundamental optical mode is zero-detuned from the mean exciton energy at normal incidence. The width of the metallic mirrors alternates, as depicted in Fig.~\ref{Fig1}. Two configurations are possible.

The {\it trivial} configuration, shown in Fig.~\ref{Fig1}(top), consists of an array of cavities where the the odd mirrors have a width of $L_{M,odd}$ , whereas the even mirrors' width equals $L_{M,even}$, with $L_{M,odd}>L_{M,even}$.
The {\it topological} array is obtained by switching the order of the mirrors. 

Intuitively, we expect that cavities separated by a thin mirror couple more efficiently than those distanced by a thicker mirror. Thus, the trivial configuration allows for the effective coupling of the cavities by pairs, as in the trivial phase of the SSH model. On the other hand, for the topological configuration, only the internal cavities of the stack couple efficiently, whereas the two cavities at the edges of the array appear as {\it isolated}, as in the topological phase of the SSH model.

Our proposal is completely general and independent of the specific organic molecule employed. However, to highlight the experimental feasibility of our setup, we illustrate our results for a concrete dye-doped polymer. Erythrosine B (ErB) has already proved as a suitable molecule for the realization of exciton-polaritons at room temperature~\cite{garcia2022fate}.

\section{SSH Polaritons: A transfer matrix-based approach}

\label{MT}
We start our theoretical study employing the transfer matrix method. This is a simple yet powerful tool to study light propagation in multilayer systems with ideal planar and parallel interfaces~\cite{yeh}.

Our setup, illustrated in Fig.~\ref{Fig1}, consists of $4N+3$ layers: $2N+1$ silver (Ag) mirrors, $2N$ active layers, and $2$ semi-infinite dielectric media at the ends of the lattice. The specific active layer used in the following corresponds to poly-vinyl alcohol (PVA) mixed with ErB. Its complex refractive index, $\tilde n_{\text{ErB}}(\omega)$, is obtained from experimental measurements \cite{garcia2022fate}, and its imaginary part, $\kappa_{\text{ErB}}(\omega)$, is shown in Fig.~\ref{FigErB}. $\kappa_{\text{ErB}}(\omega)$ presents a mean exciton at $\omega_X\approx 2.32\text{eV}$ and a secondary peak at $\omega_S\approx 2.5\text{eV}.$ Here we take $\hbar=1.$

Without loss of generality, we consider both latter media to be air with $n_{\text{Air}}=1$, noticing that the introduction of a substrate in our formalism is straightforward. The Ag complex refractive index, $\tilde n_{\text{Ag}}(\omega),$ is taken from the experimental reported values \cite{palik}.

The length of all the active layers is fixed to $L_c=140\text{nm},$ the width of all odd mirrors is $L_{\text{odd}}$, while for all even mirrors it is $L_{\text{even}}$. Plane waves propagating in each layer indexed by $l$, are described by an electric field $E_l(z)=A_le^{ik_lz}+B_le^{-ik_lz}$. Here, $l=0$ denotes the first medium (air), for the mirrors and the active layers we have $4N+2>l>0,$ where for an odd $l$ light propagates in a mirror, while for an even $l$ it propagates in an active layer. Finally, $A_l$ and $B_l$ are the amplitude coefficients for the in-coming/out-going electric fields in each medium. We write the amplitudes in vectorial form $\mathbf v_{l}=[A_l,B_l]^T$ and connect the coefficients via Maxwell equations and the appropriate boundary conditions. For $s$-polarized waves and light propagating from the $l-$th to the $l+1$-th medium we obtain $\mathcal D_l \mathbf v_l=\mathcal D_{l+1} \mathbf v_{l+1}.$  Here, the dynamical matrix $\mathcal D_l$ is given by,
\begin{gather}
\mathcal D_l= \begin{bmatrix}
 1 && 1 \\
 n_l\cos\theta_l && -n_l\cos\theta_l
 \end{bmatrix}.
\end{gather} 
Through medium $l$, the phase changes by
\begin{gather}
 \mathcal P_l=\begin{bmatrix}
 e^{i\phi_l} && 0 \\
 0 && e^{-i\phi_l}
 \end{bmatrix},
\end{gather}
where $\phi_l=k^l_{z} L_l$, $L_l$ is the length of the medium, and $k^l_{z}$ is the perpendicular component of the wave-vector of the electric field and it is given by
\begin{gather}
 k^l_{z}=\frac{\omega}{c}n_l\cos\theta_l  
\end{gather}
with $\theta_l$ the angle of incidence of the light field in the $l-$th medium measured from the $z$-axis, i.e., normal to the stack. 

It is convenient to introduce $$\mathcal M_l=\mathcal D_l \mathcal P_l  \mathcal D_l^{-1}, $$ 
to write the total transfer matrix $\mathcal T$ as following
\begin{gather}
 \mathcal T=\mathcal D_0^{-1}\left(\prod_{l=1}^{4N+1} \mathcal  M_l \right) \mathcal D'_0,
\end{gather}
with $\mathcal D_0,$ and $\mathcal D'_0,$ the interfaces matrix for the air at the ends of the array.

Finally, the reflectance can be calculated as
\begin{gather}
R=\left|\frac{ \mathcal T(2,1)}{\mathcal T(1,1)}\right|^2.
\end{gather}

\begin{figure}[h]
\centering
\includegraphics[width=\columnwidth]{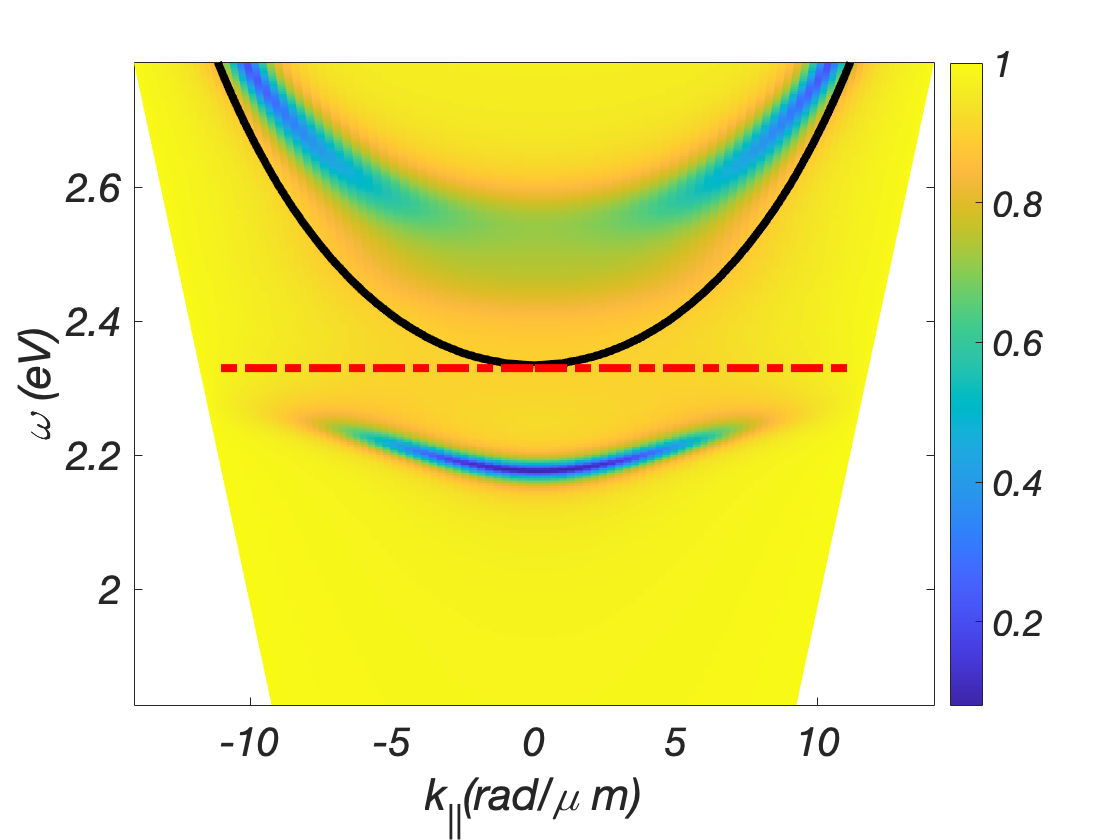}
\caption{$s$-polarized reflectance for a single cavity having $L_c=140\text{nm}$, $L_1= \text{40nm}$ and $L_3$ semi-infinite. The dashed line indicates the energy of the bare exciton peak centered around $\omega_X=2.32\text{eV},$ while the solid red line corresponds to the bare cavity photon dispersion.}
\label{Fig4}
\end{figure}

{\it Single cavity.-} Before we explore the reflectance for the two configurations shown in Fig.~\ref{Fig1}, let us remind the reflectance spectrum for a single cavity. The experimental study of Frenkel polaritons in a single cavity was subject of study in Ref.~\cite{garcia2022fate}. The reflectance spectrum for a single cavity of length $L_c=140\text{nm}$ features two polariton branches. The lower polariton arises as well-defined quasiparticle whereas the upper polariton emerges as an ill-defined polariton at small angles and only becomes a quasiparticle at large angles. As discussed in Ref.~\cite{garcia2022fate}, the blurring of the upper polariton is an inherent feature of Frenkel polaritons and dramatically influence the quasiparticle character of the polaritons. The separation between the polariton branches at normal incidence is estimated to be around $2\Omega=0.33\text{eV}.$

\begin{figure}[h]
\centering
\includegraphics[width=\columnwidth]{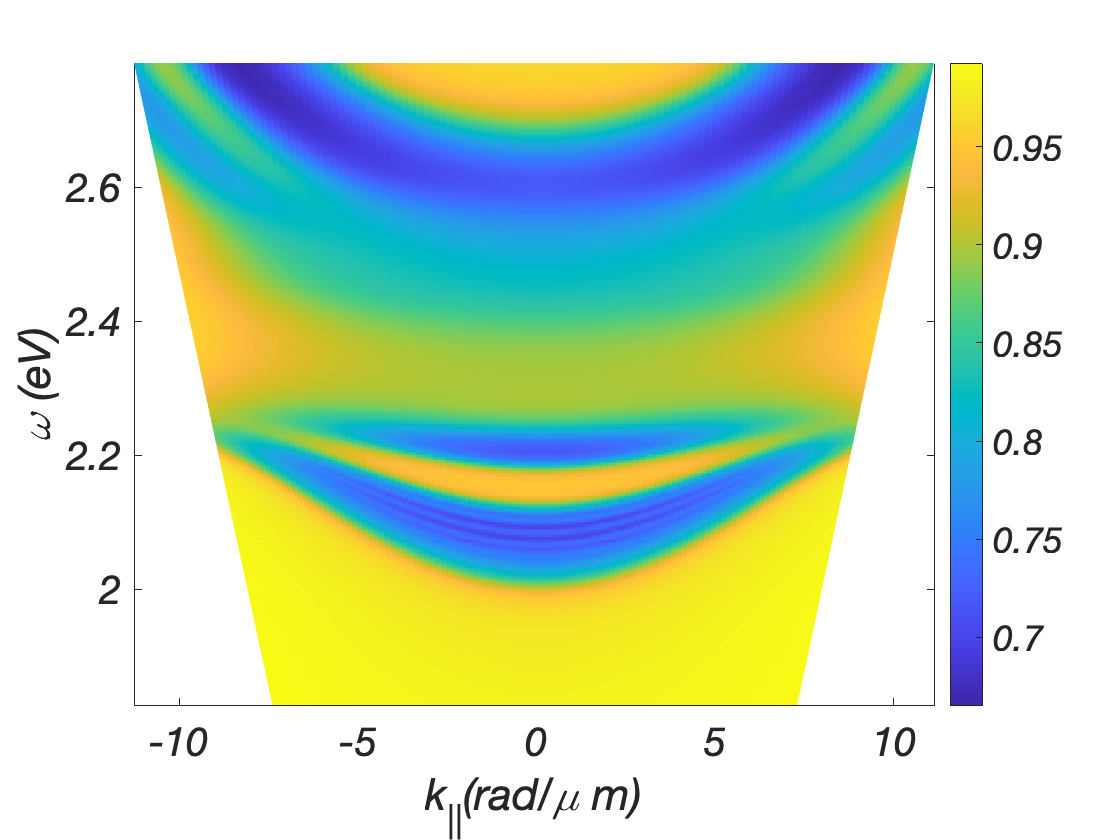}
\caption{$s$-polarized reflectance for the trivial configuration where four polariton bands appear with two bandgaps above and below the bare exciton energy.}
\label{Fig2}
\end{figure}

{\it Trivial.-} Let us start discussing the trivial configuration. We take $N=10,$ that is 20 cavities, the width of all odd mirrors is $L_{\text{odd}}=30\text{nm}$, while for all even mirror it is $L_{\text{even}}=40\text{nm}$. In Fig.~\ref{Fig2} we show the reflectance for this configuration as a function of $\omega$ and $k_{||}=\frac{\omega}{c}\sin\theta$. Below the energy of the bare exciton two polariton bands appear separated by a bandgap that becomes maximal at normal incidence and closes for large $k_{||}$ values. In the limit of infinite N, both bands form a continuum. However, for finite $N$ a slight discreteness in the bottom polariton band is expected.
Above the bare exciton energy, two upper polariton bands arise. As a consequence of the vibronic {\it shoulder} of the exciton absorption, at normal incidence only one of these bands is clearly resolved. For large $k_{||}$, the two upper polariton bands are clearly distinguishable and exhibit a bandgap. We note two facts: (i) the bandgaps opened by stacking the cavities are smaller than the splitting between the two upper and lower polariton bands; (ii) these bandgaps lie at the spectral position of the polaritons for a single cavity, shown in Fig.~\ref{Fig4}.

\begin{figure}[h]
\centering
\includegraphics[width=\columnwidth]{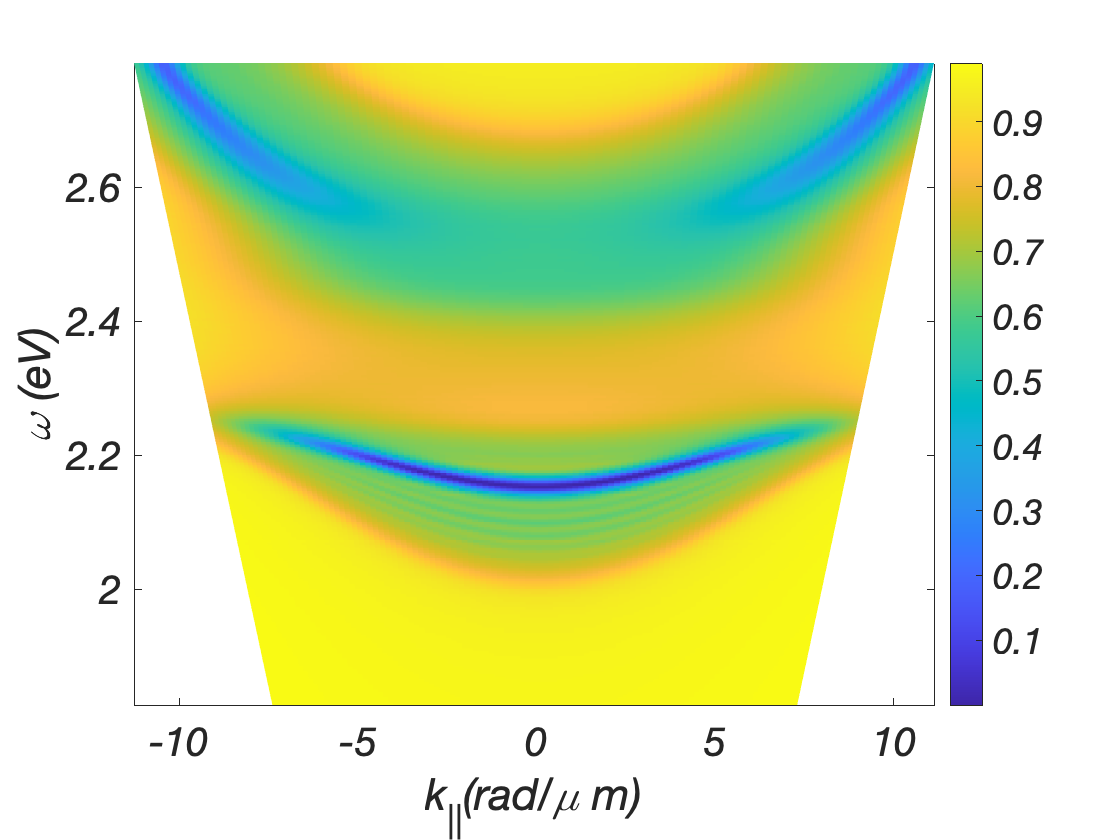}
\caption{$s$-polarized reflectance for the topological configuration. Reflectance minima arise within the two bandgaps observed for the trivial configuration. }
\label{Fig3}
\end{figure}

{\it Topological.-} We now turn our attention to the topological configuration illustrated in Fig.~\ref{Fig1}(b). The reflectance spectrum obtained from the TMM is shown in Fig.~\ref{Fig3} and exhibits striking features compared to the {\it trivial} configuration. In this case, the reflectance minima locates inside the bandgaps found for the {\it trivial} configuration and closely resembles the upper and lower polaritons of the single cavity, displayed in Fig.~\ref{Fig4}. In accordance with our previous discussion, only the lower polariton remains well-defined for all incident angles. The upper polariton state visibly blurres at normal incidence.

The spatial distribution of the normalized electric field intensity, $|E(z)/E_{Max}|^2$, is shown in Fig.~\ref{Fig7} (c) as a function of $z$ (blue curve) for the topological configuration. The electric field peaks in the odd cavities whereas significantly drops and essentially vanishes inside the even cavities. The intensity of the electric field in the odd cavities decays exponentially which further hints the topological character of our setup.

The TMM strongly suggests that our setup is analogous to the SSH model for exciton-polaritons. However, to explicitly unveil the link with the SSH model, in the following sections we develop a tight-binding model for the exciton-polaritons and contrast it to the TMM.

\section{SSH Polaritons: An effective tight-binding model approach}
\begin{figure}[h]
\centering
\includegraphics[width=0.95\columnwidth]{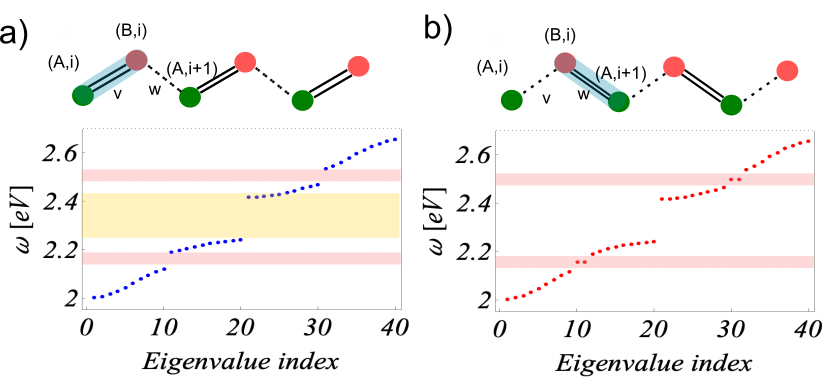}
\caption{(a) Eigenvalues for the trivial configuration at normal incidence: four polariton bands corresponding to the two branches of the lower and upper polaritons. Inset shows a cartoon of the notation employed. (b) Eigenvalues for the topological configuration, inside the energy gaps of the polariton bands two edge states per branch appear. We take $2N=20$ cavities giving $40$ eigenvalues. }
\label{Fig5}
\end{figure}

The following Hamiltonian describes a set of $2N$ coupled cavities that can be arranged either in the trivial or topological configuration, as illustrated in Fig.~\ref{Fig5}(top),
\begin{gather}\nonumber
\hat H=\sum_{i=1}^N\omega_c(\theta)\left(\hat a_{i}^\dagger\hat a_{i}+\hat b_{i}^\dagger\hat b_{i}\right)+\omega_X\left(\hat x_{i,A}^\dagger\hat x_{i,A}+\hat x_{i,B}^\dagger\hat x_{i,B}\right)\\ \nonumber
+\Omega\sum_{i=1}^N\left(\hat a_{i}^\dagger\hat x_{i,A}+\hat b_{i}^\dagger\hat x_{i,B}+\text{h.c}\right)\\ \nonumber
-\sum_{i=1}^N\left(v(\hat a_{i}^\dagger\hat b_{i}+\hat b_{i}^\dagger\hat a_{i})+w(\hat a_{i+1}^\dagger\hat b_{i}+\hat b_{i}^\dagger\hat a_{i+1})\right), \\  
\label{HP}
\end{gather}
here, $\hat a_i^\dagger$ and $\hat b_i^\dagger$ create a cavity photon in a site $A$ and $B$ respectively with energy $\omega_c(\theta)$ which depends on the incident angle $\theta$ given by the solid black line in Fig.~\ref{Fig4}. On the other hand $\hat x^\dagger_{i,A}$ and  $\hat x^\dagger_{i,B}$ create excitons with site index $i$ in the cavity $A$ y $B$, respectively. Here, the energy of the excitons is $\omega_X.$ Excitons and photons couple with a strength $\Omega$ only if all site indices are equal. Adjacent cavities couple through the tunneling of photons, where the tunneling amplitude is given either by $v$ or $w$, depending on the configuration, as illustrated in Fig.~\ref{Fig5} (top).

We now study the tight-binding model for the exciton-polaritons within the SSH model. For reasons will become clear later we take hopping coefficients of $v=0.15$ and $w=0.09$ for the trivial configuration, whereas for the topological we simply swap these coefficients, i.e., $w=0.15$ and $v=0.09.$ The SSH model for exciton-polaritons is a simple quadratic Hamiltonian that can straightforwardly be diagonalized. For consistency with the TMM we take $N=10$ corresponding to 20 cavities.
\begin{figure*}[ht!]
\centering
\includegraphics[width=2\columnwidth]{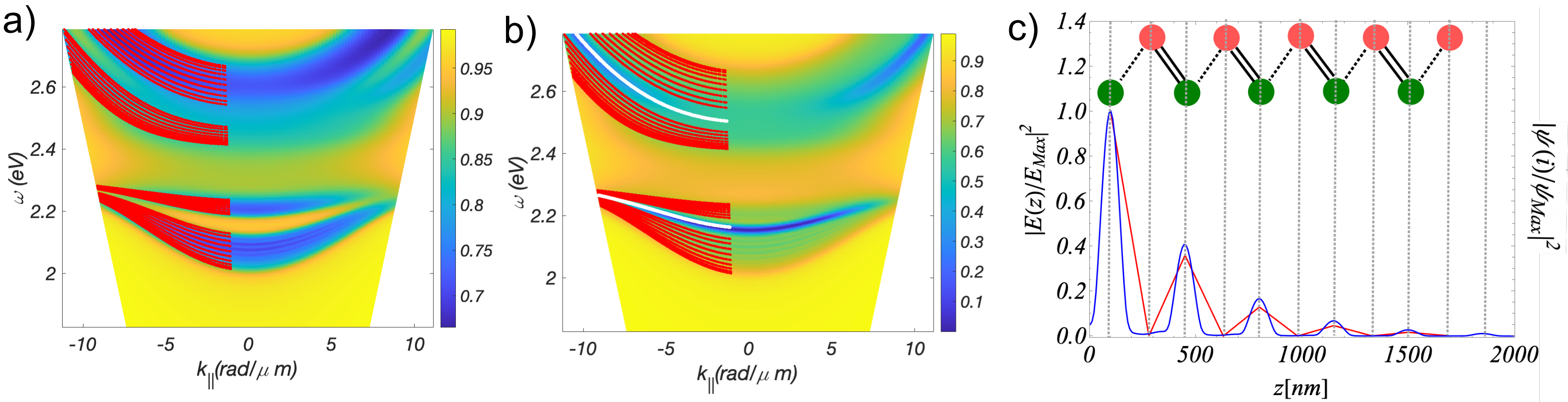}
\caption{(a-b)Eigenvalues of the SSH model in Eq.~\ref{HP} as a function of the in-plane momentum  $k_{||}$ illustrated by the red dots. The energies of the SSH model are plotted on top of the reflectance spectrum for (a) trivial configuration and (b) topological configuration. The white dots are the eigenvalues of Eq.~\ref{HP} corresponding to edge states and lie at the energies of the bare lower and upper polaritons. (c) Normalized electric field intensity as a function of $z,$ (blue curve). Amplitude of the wavefunction obtained from the SSH model (red curve). The gray vertical lines give the position of the center of the cavities.}
\label{Fig7}
\end{figure*}

{\it Trivial.-} We start discussing the trivial configuration. For clarity, we show in Fig.~\ref{Fig5}(a) the eigenvalues of the Hamiltonian in Eq.~\ref{HP} considering first normal incidence and resonant conditions $\omega_c(\theta=0)=\omega_X$. In this case we observe that the lower and upper polariton states split leading to four polariton bands: two above and two below the energy of the bare exciton energy. The lower and upper polaritons yield to two bands separated by a gap, marked by the pink area. The Rabi coupling leads to the avoided crossing (yellow area) that separates the lower from the upper polariton bands. These four polariton bands display a difference in their bandwidths. Specifically, the lowest polariton band is broader than the second polariton band. This feature, also visible in the TMM calculations, can hardly be explained within the TMM. Conversely, the tight-binding model provides a very intuitive physical explanation for it. When photons hybridise with excitons forming polaritons, the photon tunneling between adjacent cavities depends on the polaritons Hopfield coefficents, i.e., the coupling efficiency of polaritons living in adjacent cavities depends on their photonic/excitonic component. Thus, one expects that polaritons with large excitonic component exhibit a weak tunnelling leading to narrow polariton bands. This is the case for the two bands located around the bare exciton energy. On the other hand, polaritons with large photonic component lead to a broad bandwidth consequence of dispersion and enhanced hopping.

Formally, these arguments can be read  by studying the hopping terms of the Hamiltonian. For instance, the tunneling between photons $A$ and $B$ with same site index, $i$, in the polariton basis ($\hat L_{i,\alpha},\hat U_{i,\alpha})$ with $\alpha=A, B$ is 
\begin{gather} \nonumber
-v\left(\hat a_{i}^\dagger\hat b_{i}+\text{h.c}\right)= -v\left(\mathcal S_{i,A}\mathcal S_{i,B}\hat L^\dagger_{i,A}\hat L_{i,B}\right.\\ \nonumber+\mathcal C_{i,A}\mathcal C_{i,B}\hat U_{i,A}^\dagger\hat U_{i,B} 
+\mathcal S_{i,A}\mathcal C_{i,B}\hat L^\dagger_{i,A}\hat U_{i,B}\\ \nonumber
\left.+\mathcal C_{i,A}\mathcal S_{i,B}\hat U^\dagger_{i,A}\hat L_{i,B}+\text{h.c}\right).\\
\label{EqTR}
\end{gather}
Here, the photon written in the basis of the lower and upper polariton in terms of the standard Hopfield coefficients is $\hat a_i/\hat b_i=\mathcal S_{i,\alpha}\hat L_{i,\alpha}+\mathcal C_{i,\alpha}\hat U_{i,\alpha},$
 with $\mathcal S_{i,\alpha}^2+\mathcal C^2_{i,\alpha}=1,$ where $$\mathcal C_{i,\alpha}^2=\frac{1}{2}\left(1+\frac{\omega_c(\theta)-\omega_X}{(\omega_c(\theta)-\omega_X)^2+4\Omega^2}\right).$$ Equation~\ref{EqTR} stresses that adjacent cavity polaritons can only couple via their photonic component. Thus, states with very small photonic components have suppressed tunnelling and tend to localize within the corresponding cavities. Such localization corresponds to the band flattening observed for the two bands that are close to the bare exciton energy (Fig.~\ref{Fig5}). In contrast, for states with a large photonic component, the tunnelling of polaritons becomes essentially the bare photon term, $v,$ which makes the two polariton bands far detuned from the exciton dispersive and broad.

We remark that the understanding of the narrowing of the bands close to the {\it bare} exciton line cannot straightforwardly be read from the TMM. This discussion naturally arises from the tight-binding formalism providing a deeper insight into the physical setup. 

We can further understand the tight-binding model and its equivalence to the experimental proposal if we vary the incident angle $\theta$. In Fig.~\ref{Fig7} (a) we plot the eigenvalues of the Hamiltonian (red dots) in Eq.~\ref{HP} as a function of $k_{||}.$ For clarity in our comparison we show in the background the reflectance spectrum obtained from the TMM. The remarkable agreement between the TMM and the SSH model for exciton-polaritons support our experimental proposal. Furthermore, we also observe the closing of the lower bandgap as the lower polariton becomes more excitonic at larger incident angles, in clear agreement with our previous discussion based on the Hopfield coefficients.

{\it Topological.-} The topological configuration at normal incidence yields to the eigenvalues in Fig.~\ref{Fig5} (b). In addition to the four polariton bands, we observe the appearance of two edge states located in the middle of the polariton gaps (pink area) whose energy lies exactly at the energy of the polaritons sustained by the single cavity:
\begin{gather}
 \omega_{\text{LP/UP}}(\theta)=\frac{1}{2}\left(\omega_c(\theta)+\omega_X\mp\sqrt{(\omega_c-\omega_X)^2+4\Omega^2}\right).  
\end{gather}
Since we consider resonant conditions, in Fig.~\ref{Fig5} (b) the edge states are distanced precisely by  $2\Omega.$

For varying angle of incidence, we observe in Fig.~\ref{Fig7} (b) the persistence of these edge states which remain confined within the corresponding bandgaps. For clarity, we have highlighted the energy of these states with white markers whereas the red dots correspond to the bulk states. In accordance with our previous analysis, also for the topological configuration the bandgap associated with the lower polariton bands closes for large angles. This is consequence of the large excitonic component of the polaritons. On the other hand, the gap separating the two upper bands slightly increases at large angles as polaritons become predominantly photonic.

Finally, in Fig.~\ref{Fig7} (c) we show the distribution of the wavefunction for the edge state along the cavities. The wavefunction is only non-zero for the $A$ cavities whereas it vanishes for the $B$ cavities. The amplitude of the wavefunction in the $A$ cavities decays exponentially with the index site $i.$   At each site, the state of the polariton retains the maximal coupling between the excitons and photons, that is, the Hopfield coefficients equal to 1/2. The distribution of the amplitude of the wavefunction predicted by the tight-binding model for exciton-polaritons agrees remarkably well with the electric field intensity obtained with the TMM. It captures both the vanishing of the light intensity for the $B$ cavities and the exponential decay observed in the $A$ cavities as a function of $z$. Note that the electric field is a continuous function of $z$ where the wavefunction is discrete in the index site $i.$

The SSH model for exciton-polaritons has added remarkable physical insights to the polariton physics predicted by the TMM. However, features that extend beyond the single-particle approach of polaritons are beyond the realm of the SSH model, hence, cannot be captured by this model. For instance, the breakdown of the upper polariton  at normal incidence which washes out the bandgap of the upper bands cannot be obtained within our SSH model for polaritons. This remarks the need for the dual approach: on the one hand, the TMM absent of any fitting parameters provides a powerful tool that gives the reflectance spectrum that should be experimentally observed but does not link directly to the SSH model. On the other hand, the tight-binding model allows us to link the phenomenology of the TMM with the SSH model and provides deep physical insight, yet it fails to contain the full complexity of the system. By combining these two approaches we obtain the complete picture of the SSH exciton-polaritons both from a pragmatic experimental point of view and the fundamental understanding of the model.

\section{Experimental Considerations: Robustness to Fabrication imperfections}

In practice, there are experimental considerations that may limit the realization of the SSH array of polaritons that require discussion. 
First, incoherent processes such as photon leaking, non-radiative losses of the excitons, and coupling to additional excitonic modes. Second, the ability to produce mirrors with uniform widths and cavities with different lengths. Finally, limitations to realize a large number of cavities, that is, finite size effects.

In our TMM formalism, we have included the experimental values of the refractive index of the active layer, this includes all of the incoherent matter processes. On the other hand, the leaking of the photons is naturally included and arises as a broadening of the photonic lines. Our results discussed previously demonstrate that the SSH for exciton-polaritons is very robust towards these effects. The topological effects for the lower polariton bands are well-defined at all incident angles. For the upper polaritons, the breaking of the quasiparticle picture leads to a blurred region where the edge modes are hardly visible, however, with the opening of the angle these states become well-defined. In both cases, we have found a very good agreement with the SSH polaritons treated at the single-particle level.

To study the effects of fabrication imperfections, we add a random and different error to all of the widths of both the mirrors and cavities. Experimentally, we estimate that the mirrors and cavities can be realized within an error of circa $\text{4nm},$ thus we add an error for the fabrication of the mirrors of $5\%,$  that is, we take $L^i_{M,\text{even/odd}}=L_{M,\text{even/odd}}^0+\delta L^i_{M,\text{even,odd}},$ where $L_{M,even/odd}^0$ is the length of the mirrors discussed previously and $\delta L^i_{M,\text{even,odd}}$ a random number taken different for each site $i.$ We also consider an error  for the cavity lengths
 $L^c_i=L^c_0+\delta L_i$ with $L^0_c=140\text{nm}$ and an error of  $\delta L_i$ in between the range $(-3.5,3.5)\text{nm},$ again, different for each cavity. 
 
The reflectance spectrum adding these fabrication imperfections for an array of 20 cavities is shown in Fig.~\ref{Fig9} (a) for the topological configuration. We obtain a reflectance that closely resembles the uniform case in Fig.~\ref{Fig3}. This remarks  that our proposal is indeed robust to a defects on the experimental procedure that may produce mirrors and cavities with small errors in their widths.

Finally, we study the reflectance spectrum for a set of eight cavities, that is $N=4$ dimers, here we also retain  the fabrication imperfections discussed above together with the experimental absorption spectrum of the organic molecules. The reflectance is  shown in Fig.~\ref{Fig9}(b), we observe clearly the edge mode of the lower lower band, whereas the edge mode of the upper band becomes distinguishable at large angles. Note, that Fig.~\ref{Fig9}(b) captures all the relevant features of Fig.~\ref{Fig3} and Fig.~\ref{Fig7}(b).
\begin{figure}[H]
\centering
\includegraphics[width=0.47\columnwidth]{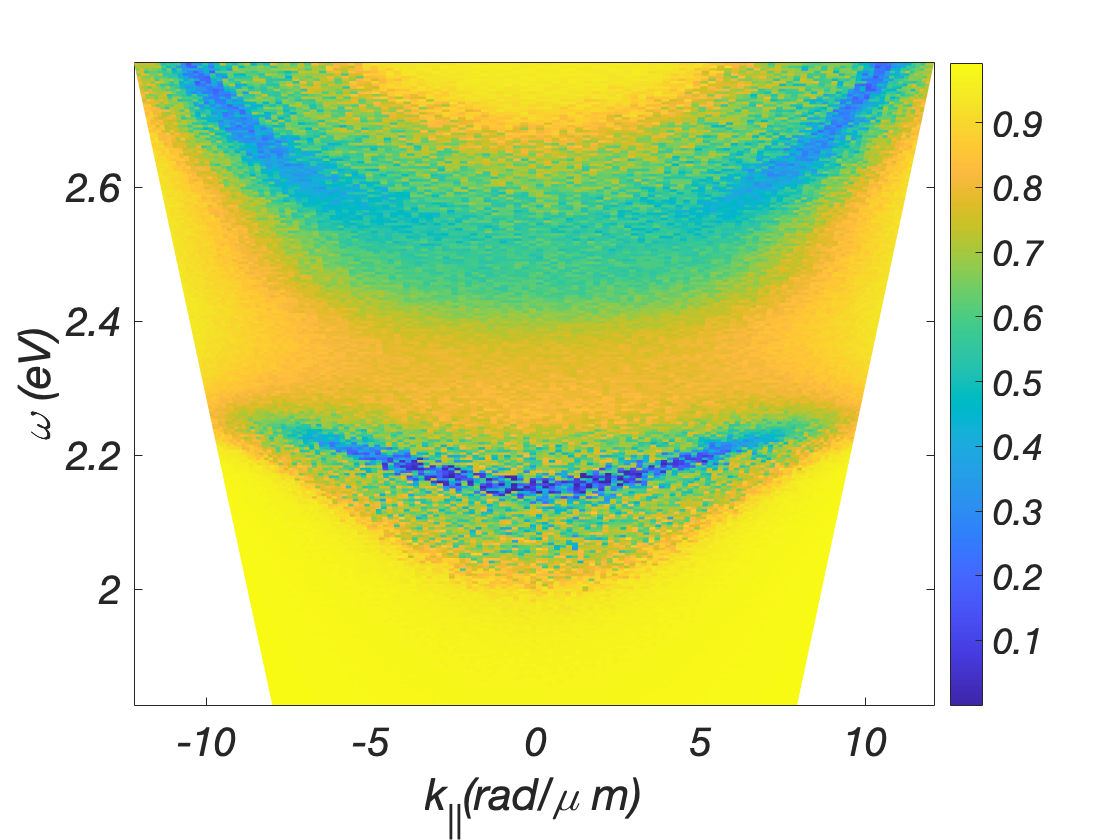}
\includegraphics[width=0.47\columnwidth]{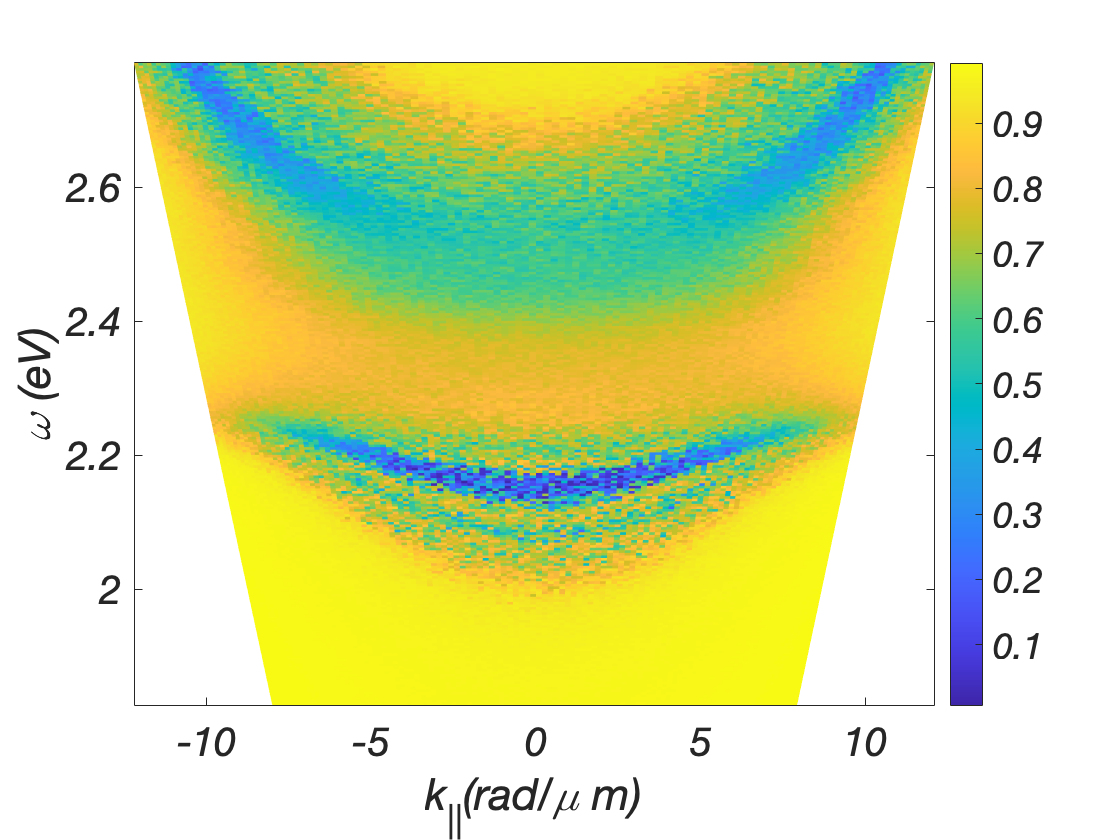}
\caption{Reflectance spectrum in the presence of imperfections for (a) $N=10$ and (b) $N=4,$ that is for $20$ and $8$ cavities respectively for the topological configuration. }
\label{Fig9}
\end{figure}

Our findings allow us to conclude that our experimental proposal is robust to the underlying complexity of the exciton spectrum, inherent fabrication errors and limited number of cavities. Therefore, stands as a promising platform to study the SSH model for organic polaritons at room temperature.

\section{Ideal Excitons}
Let us turn our attention to the study of ideal excitons. Here, the absorption spectrum of the active layer is single peaked and the incoherent processes coming from the vibronic coupling are removed. This scenario is more commonly found in inorganic materials. The imaginary part of the refractive index is shown in Fig.~\ref{Fig1} with the dashed blue curve and it consists of a single narrow peak centered around $\omega\approx 2.32\text{eV}$ with an oscillatory strength of $2\Omega=0.33\text{eV}$ and a small exciton broadening of $\gamma_X=0.025\text{eV}$. 
\begin{figure}[H]
\centering
\includegraphics[width=0.47\columnwidth]{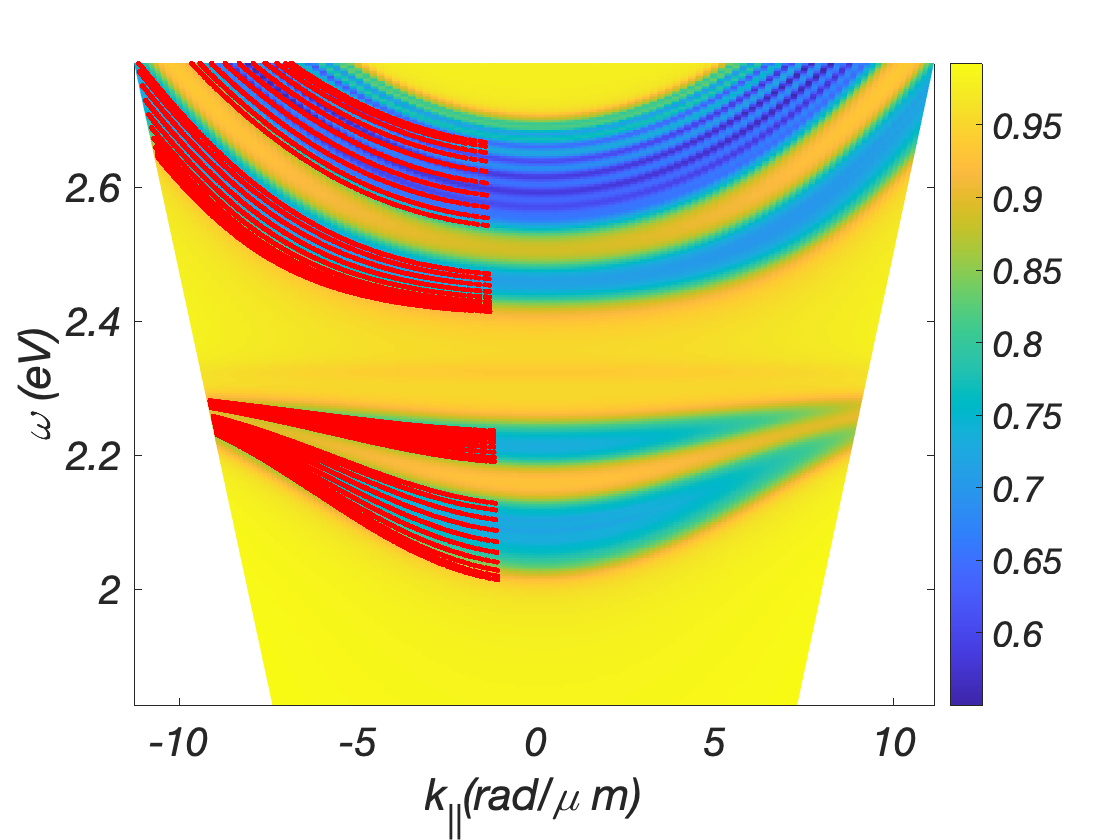}
\includegraphics[width=0.47\columnwidth]{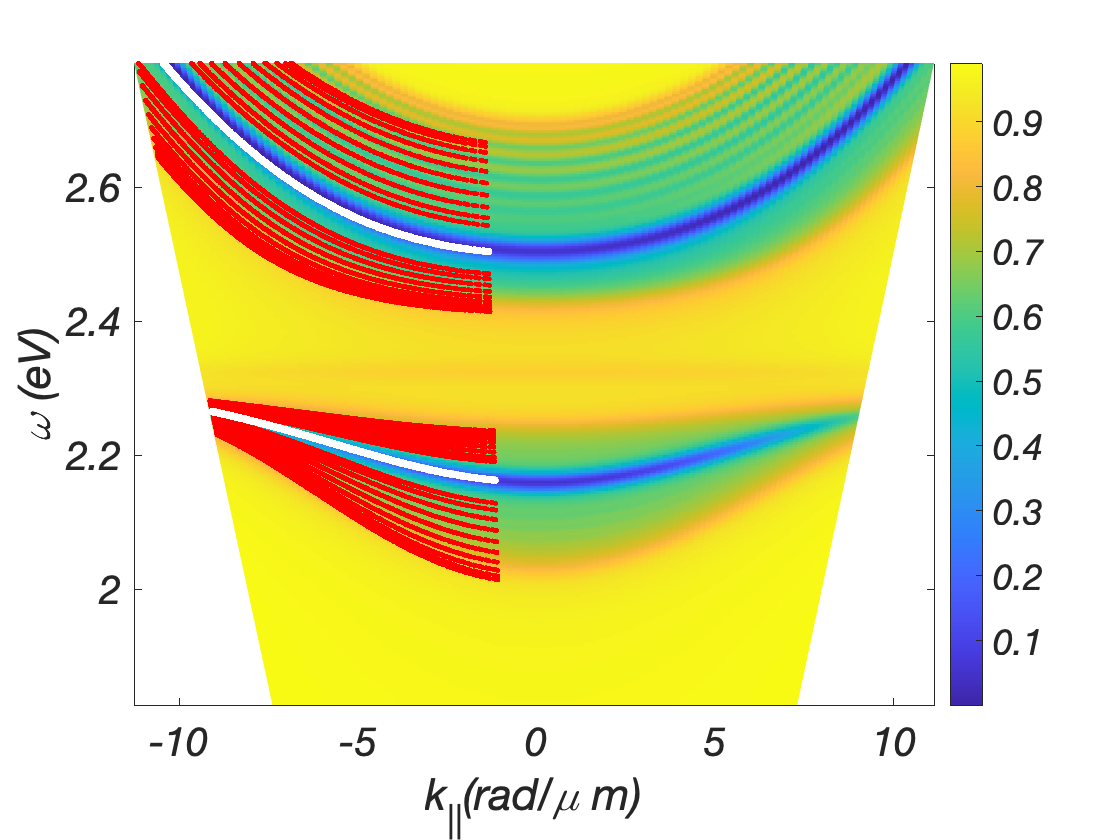}
\caption{s-polarized reflectance for the one-dimensional lattice of cavities containing a material with ideal excitonic response. (a) Trivial and (b) topological configuration. }
\label{Fig9}
\end{figure}
In Fig.~\ref{Fig9} we calculate the $s$-polarized reflectance for the trivial and topological configurations of the one-dimensional lattice where the organic material has been replaced by one with lorentzian excitonic response. Figure~\ref{Fig9}(a) corresponds to the trivial configuration and closely resembles Fig.~\ref{Fig7}(a). However, in this case, the upper branches are well-defined even at normal incidence and the four polariton branches are clearly visible. As expected, in the absence of the matter incoherent processes, the two edge states existing in the topological configuration appear well-defined for all incident angles, as shown in Fig.~\ref{Fig9}(b).

\section{Perspectives and Conclusions}
Frenkel polaritons offer a tunable platform to realize topological phases of light and matter. The ability to produce topological states at ambient conditions is a necessary condition to deliver their promise on technological applications such as integrated quantum optical circuits, non-linear light, and chiral and topological lasers.

In this article, we have studied a one-dimensional lattice of nanocavities filled with a dye-doped polymer strongly coupled to light. By using two complementary approaches, we have demonstrated the direct analogy between the polariton band structure of the lattice to the one-dimensional SSH model. First, we have calculated the propagation of the light field across the lattice by using the transfer matrix method. The spectra strongly depend on the configuration of the lattice: in the {\it trivial} phase we observed four polariton bands, two lower bands below the exciton energy separated by a bandgap, and two upper bands above the exciton energy also distance by a bandgap; conversely, in the topological phase, we obtained two polariton states whose dispersion falls within the bandgaps and whose electric field intensity localizes around the edge cavity, exponentially decaying within the lattice.

We complemented our analysis with an effective tight-binding model, which allows us to link the reflectance spectra with the SSH model. By combining these approaches we obtained a comprehensive understanding both from the experimentally relevant picture and with the elementary blocks of the single-particle polariton topological physics. Our works provides valuable benchmark for future theories on lattices of Frenkel polaritons and realistic experimental implementations of the SSH model for Frenkel polaritons at ambient conditions.

\section{Acknowledgments} G. P. acknowledges financial support from Grants UNAM DGAPA PAPIIT No. IN104522 and CONACyT projects 1564464 and 1098652. H. L. G. acknowledges financial support from Grant UNAM DGAPA PAPIIT No. IA103621. A. C. G. acknowledges financial support from Grant UNAM DGAPA PAPIIT No. IN108620. C. L. O-R acknowledges financial support from Grant UNAM DGAPA PAPIIT IG100521.
\bibliography{references}
\end{document}